\documentclass{aastex631}

\usepackage{array}
\usepackage{csquotes}
\newcolumntype{L}{>{\centering\arraybackslash}m{8cm}}
\newcolumntype{Q}{>{\centering\arraybackslash}m{4cm}}

\begin{document}

\title{The NASA Exoplanet Archive and Exoplanet Follow-up Observing Program: Data, Tools, and Usage}

\correspondingauthor{Jessie L. Christiansen}
\email{christia@ipac.caltech.edu}
\author[0000-0002-8035-4778]{Jessie L. Christiansen}
\affiliation{NASA Exoplanet Science Institute, IPAC, MS 100-22, Caltech, 1200 E. California Blvd, Pasadena, CA 91125}

\author{Douglas L. McElroy}
\affiliation{NASA Exoplanet Science Institute, IPAC, MS 100-22, Caltech, 1200 E. California Blvd, Pasadena, CA 91125}

\author{Marcy Harbut}
\affiliation{NASA Exoplanet Science Institute, IPAC, MS 100-22, Caltech, 1200 E. California Blvd, Pasadena, CA 91125}

\author[0000-0002-5741-3047]{David R. Ciardi}
\affiliation{NASA Exoplanet Science Institute, IPAC, MS 100-22, Caltech, 1200 E. California Blvd, Pasadena, CA 91125}

\author{Megan Crane}
\affiliation{NASA Exoplanet Science Institute, IPAC, MS 100-22, Caltech, 1200 E. California Blvd, Pasadena, CA 91125}

\author{John Good}
\affiliation{NASA Exoplanet Science Institute, IPAC, MS 100-22, Caltech, 1200 E. California Blvd, Pasadena, CA 91125}

\author[0000-0003-3702-0382]{Kevin K. Hardegree-Ullman}
\affiliation{NASA Exoplanet Science Institute, IPAC, MS 100-22, Caltech, 1200 E. California Blvd, Pasadena, CA 91125}

\author[0000-0002-3239-5989]{Aurora Y. Kesseli}
\affiliation{NASA Exoplanet Science Institute, IPAC, MS 100-22, Caltech, 1200 E. California Blvd, Pasadena, CA 91125}

\author[0000-0003-2527-1598]{Michael B. Lund}
\affiliation{NASA Exoplanet Science Institute, IPAC, MS 100-22, Caltech, 1200 E. California Blvd, Pasadena, CA 91125}

\author{Meca Lynn}
\affiliation{NASA Exoplanet Science Institute, IPAC, MS 100-22, Caltech, 1200 E. California Blvd, Pasadena, CA 91125}

\author{Ananda Muthiar}
\affiliation{NASA Exoplanet Science Institute, IPAC, MS 100-22, Caltech, 1200 E. California Blvd, Pasadena, CA 91125}

\author[0000-0002-5408-4954]{Ricky Nilsson}
\affiliation{NASA Exoplanet Science Institute, IPAC, MS 100-22, Caltech, 1200 E. California Blvd, Pasadena, CA 91125}

\author{Toba Oluyide}
\affiliation{NASA Exoplanet Science Institute, IPAC, MS 100-22, Caltech, 1200 E. California Blvd, Pasadena, CA 91125}

\author{Michael Papin}
\affiliation{NASA Exoplanet Science Institute, IPAC, MS 100-22, Caltech, 1200 E. California Blvd, Pasadena, CA 91125}

\author{Amalia Rivera}
\affiliation{NASA Exoplanet Science Institute, IPAC, MS 100-22, Caltech, 1200 E. California Blvd, Pasadena, CA 91125}

\author{Melanie Swain}
\affiliation{NASA Exoplanet Science Institute, IPAC, MS 100-22, Caltech, 1200 E. California Blvd, Pasadena, CA 91125}

\author{Nicholas D. Susemiehl}
\affiliation{NASA Exoplanet Science Institute, IPAC, MS 100-22, Caltech, 1200 E. California Blvd, Pasadena, CA 91125}

\author{Raymond Tam}
\affiliation{NASA Exoplanet Science Institute, IPAC, MS 100-22, Caltech, 1200 E. California Blvd, Pasadena, CA 91125}

\author[0000-0003-2192-5371]{Julian van Eyken}
\affiliation{NASA Exoplanet Science Institute, IPAC, MS 100-22, Caltech, 1200 E. California Blvd, Pasadena, CA 91125}

\author[0000-0002-5627-5471]{Charles Beichman}
\affiliation{NASA Exoplanet Science Institute, IPAC, MS 100-22, Caltech, 1200 E. California Blvd, Pasadena, CA 91125}
\affiliation{Jet Propulsion Laboratory, California Institute of Technology, Pasadena, CA 91109}



\begin{abstract}

The NASA Exoplanet Archive and the Exoplanet Follow-up Observing Program service are two widely used resources for the exoplanet community. The NASA Exoplanet Archive provides a complete and accurate accounting of exoplanetary systems published by NASA missions and by the community in the refereed literature. In anticipation of continued exponential growth in the number of exoplanetary systems, and the increasing complexity in our characterization of these systems, the NASA Exoplanet Archive has restructured its primary tables and interfaces, as well as extending and standardizing their modes of access. The Exoplanet Follow-up Observing Program service provides the exoplanet community with a venue for coordinating and sharing follow-up and precursor data for exoplanets, their host stars, and stars that might eventually be targets for future planet searches, and recently reached one million files uploaded by the community. In this paper we describe the updates to our data holdings, functionality, accessibility, and tools, as well as future priorities for these two services.

\end{abstract}

\keywords{Astronomy databases -- Exoplanet catalogs -- Exoplanets}


\section{Introduction} \label{sec:intro}

After decades of accelerating interest, exoplanets are now at the forefront of astrophysics. Comprising a primary scientific focus of NASA’s next flagship mission recommended for study by the Decadal Survey on Astronomy and Astrophysics 2020: Pathways to Discovery in Astronomy and Astrophysics for the 2020s (Astro2020),\footnote{Astro2020: \url{https://www.nationalacademies.org/our-work/decadal-survey-on-astronomy-and-astrophysics-2020-astro2020}} $\sim$30\% of the time allocated on NASA’s James Webb Space Telescope (JWST) in Cycles 2 and 3,\footnote{JWST Cycle 2 Results: \url{https://www.stsci.edu/files/live/sites/www/files/home/jwst/science-planning/user-committees/jwst-users-committee/_documents/jwst-cycle2-
peer-review-results.pdf}} and an ever-increasing number of refereed papers, exoplanet science is well established as a productive field with a growing base of practitioners.

The scale and complexity of exoplanet science have increased significantly since the previous paper describing the NASA Exoplanet Archive (NEA) \citep{Akeson2013}. That paper outlined the NEA data holdings as of 2013, primarily the table of confirmed exoplanets (containing fewer than a thousand planets at the time), and the high-level Kepler mission science products for which the NEA is the mission archive, as well as the first generation of stand-alone tools. Since then, the number of transiting planets discovered has risen dramatically. From NASA, the Kepler mission \citep{Borucki2010} published thousands of planets and was extended into the K2 mission \citep{Howell14}, and TESS \citep[Transiting Exoplanet Survey Satellite;][]{Ricker2015} was launched and has been extended twice, with a proposal for a third extension submitted. From the European Space Agency (ESA), the CHEOPS mission \citep[CHaracterising ExOPlanet Satellite;][]{Benz2021} is currently monitoring transiting planets; the Gaia mission \citep{GaiaC2016, GaiaC2018, GaiaC2023} has completed its observations and is expected to deliver tens of thousands of planets discovered with astrometry, and PLATO (PLAnetary Transits and Oscillations of stars), scheduled for launch in 2026, is designed to find Earth-sized transiting planets around bright nearby solar-like stars. From the ground, a new generation of [extreme] precision radial velocity ([E]PRV) instruments came online \citep[e.g.][]{KPF2016,EXPRES2016,MAROONX2018}, and both new follow-up and discovery surveys were initiated. Advancements in microlensing and direct imaging techniques from the ground have led to hundreds of discoveries by those methods---building on this success, the Nancy Grace Roman mission, scheduled for launch in 2027, will perform the first microlensing survey from space to determine the frequency of planets in outer solar systems, and will deploy a space-based high contrast imaging coronograph with precision wavefront control in a key step toward directly imaging an Earth-like planet with a mission like Habitable Worlds Observatory (HWO). Finally, the launch of NASA's JWST mission \citep{JWST2006} has led to the unprecedented ability to characterize exoplanet atmospheres of known exoplanets. All of this is reflected in the rapid growth of exoplanet research published by the community, expanding from $\sim 800$ papers per year in 2010, one year after the Kepler launch, to 2,000 papers per year in 2022.

\begin{figure}[t!]
\centering
\includegraphics[width=0.8\textwidth]{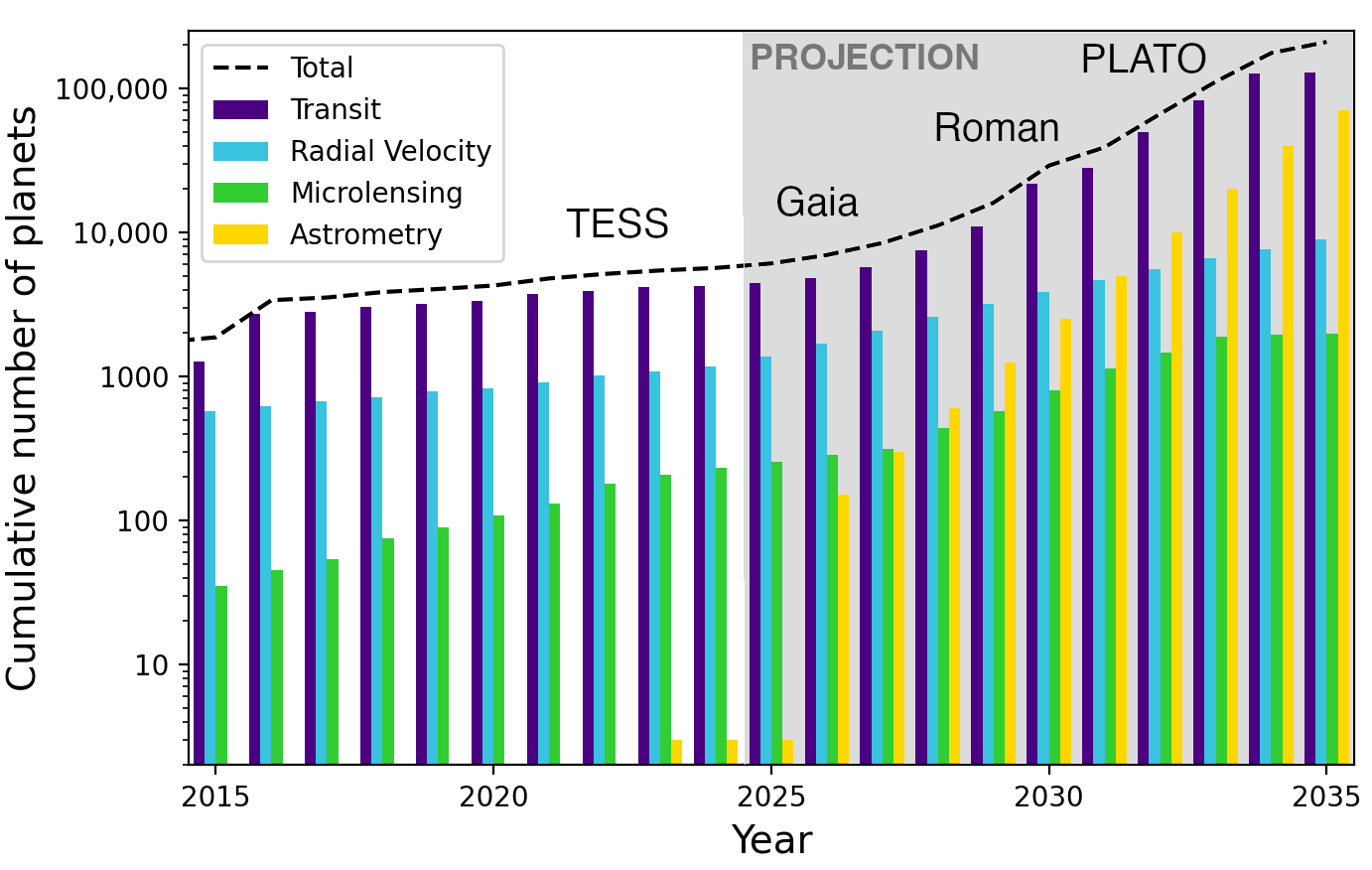}
\caption{The number of planets has been growing exponentially for decades. Here we show the decade leading up to 2025, and a projection of the expected yield of exoplanets for the following decade (see Sec. \ref{sec:literature} for a discussion of the future landscape).
\label{fig:counts}}
\end{figure}

In March 2022, the NASA Exoplanet Archive reached 5,000 confirmed planets---and is poised to serve more than 6,000 confirmed exoplanets within the year, continuing the exponential rise in discoveries over the last three decades (Figure \ref{fig:counts}). The growing field has also explored new techniques and software algorithms, producing new discoveries and previously unattainable planetary parameters. The NEA\footnote{NASA Exoplanet Archive: \url{https://exoplanetarchive.ipac.caltech.edu}} and ExoFOP\footnote{ExoFOP: \url{https://exofop.ipac.caltech.edu}} services have both evolved significantly to meet these increased needs, and as such this paper is presented to describe our current data and tools for the community. 

This paper updates and expands on \citet{Akeson2013} by outlining the latest data holdings and functionality provided by the NEA and ExoFOP, and is organized as follows: in Sec. \ref{sec:data}, we describe the current data holdings in the NEA and ExoFOP, including NASA mission data, published exoplanet detection and atmosphere data, and contributed datasets, as well as methods for accessing these data; in Sec. \ref{sec:tools} we describe the suite of tools we provide for accessing, interacting with, visualizing and fitting data in our holdings; in Sec. \ref{sec:user} we describe the processes by which we receive input from and provide support to the community; in Sec. \ref{sec:future} we describe the upcoming exoplanet landscape in the next decade and our plans for evolving our data holdings, access, and tools to support the changing community needs; and finally, in Sec. \ref{sec:acks} we provide guidance on acknowledging the services described in this paper. 

\begin{deluxetable*}{lrrr}
\tablecaption{Summary of NASA Exoplanet Archive data holdings with time. \label{tab:holdings}}
\tablehead{\colhead{Parameter} & \colhead{2015} & \colhead{2019} & \colhead{2024}}
\startdata
Confirmed exoplanets    & 1,819  & 4,019 & 5,811\\
NASA mission candidates    & 2,019  & 4,330 & 7,662 \\
Planetary solutions     & $\sim$5,000  & $\sim$26,000 &  $\sim$36,000\\
Planet parameter values     & $\sim$16,400  & $\sim$230,000 & $\sim$590,000 \\
Stellar parameter values     &  $\sim$14,000  & $\sim$130,000 & $\sim$3,670,000 \\
Peer-reviewed references     & $\sim$1,700  & $\sim$2,500 & $\sim$3,400 \\
\enddata
\end{deluxetable*}

\section{Data Holdings} \label{sec:data}
The data holdings for the NEA and ExoFOP are continuously evolving and changing with weekly (NEA) and daily (ExoFOP) updates. The data overview presented here is intended to give a summary of the data content at the time of this writing, and to give context to the scale and complexity of the data made available through the services.

\subsection{NASA Exoplanet Archive Data Holdings}
\subsubsection{Planetary System Data}
\label{sec:psdata}

The NEA’s highest priority is the core mandate of maintaining and serving a complete, up-to-date, highly accurate list of confirmed planets drawn from the refereed literature, and of NASA mission planet candidates. Criteria for inclusion of confirmed planets are detailed at \url{https://exoplanetarchive.ipac.caltech.edu/docs/exoplanet_criteria.html}. They include:
\begin{itemize}
    \item The mass (or minimum mass) of the object is equal to or less than 30 Jupiter masses.
    \item The object is not free floating.
    \item Sufficient follow-up observations and validation have been undertaken to deem the possibility of the object being a false positive as unlikely.
    \item The above information, along with further orbital and/or physical properties, are available in peer-reviewed publications.
\end{itemize}

Table \ref{tab:holdings} summarizes our current holdings, as of December 2024. Since \cite{Akeson2013} the number of exoplanets has increased six-fold, from 913 to 5,811, with over 3,900 of these from NASA’s TESS, K2 and Kepler missions. Across all confirmed planets, we added $\sim$30,000 new parameter sets from the published literature from 1,300 new references.

Each day, new and updated refereed papers are collected from the Astrophysics Data Service (ADS) and arXiv and ranked in terms of their relevance to the NEA by a machine-learning classifier, trained on previously manually selected papers (N. Susemiehl et al., in prep). NEA staff evaluate new published parameter sets against the existing holdings. If the new data extend (e.g., by adding parameters for which we do not currently have a value) or improve upon (by providing more precise measurements of existing values, and/or updated ephemerides) the existing holdings, the NEA ingests the data set. Typically, a new data set will fulfill one of these criteria, but there are publications that do not and are not ingested. For each planet, the NEA staff then selects the most complete and precise of the available parameter sets as the ``default'' parameter set. Providing multiple published parameter sets for each planet, where available, enables the community to understand the diversity of parameter sets and to choose the set appropriate for their science goals.

Our planetary systems data are served in multiple ways, depending on the needs of the users. The interactive Planetary Systems\footnote{Planetary Systems table: \url{https://exoplanetarchive.ipac.caltech.edu/table/PS}} table lists all ingested NEA parameter sets for all confirmed planets, with over 38,000 rows for the $\sim$5,900 planets (see Sec. \ref{sec:icetables} for a discussion of the functionality of our interactive tables). The table contains both the planetary parameters and stellar parameters as they were published in the literature---one row per publication per planetary solution, with the associated stellar parameters for that planetary set of parameters. 

To facilitate a more holistic view of planetary systems, the interactive Planetary Systems Composite Data\footnote{Planetary Systems Composite Data table: \url{https://exoplanetarchive.ipac.caltech.edu/table/PSCompPars}} table collapses those 36,000 parameter sets down to a single set for each planet, by filling in the gaps in the default parameter set with values with the most precise alternate value in our holdings, or calculations from published relations. This provides a maximally “filled out” parameter set for each confirmed planet, but users are cautioned the table will mix values from multiple sources where the default parameter set is not complete.

\begin{figure}[t!]
\includegraphics[width=\textwidth]{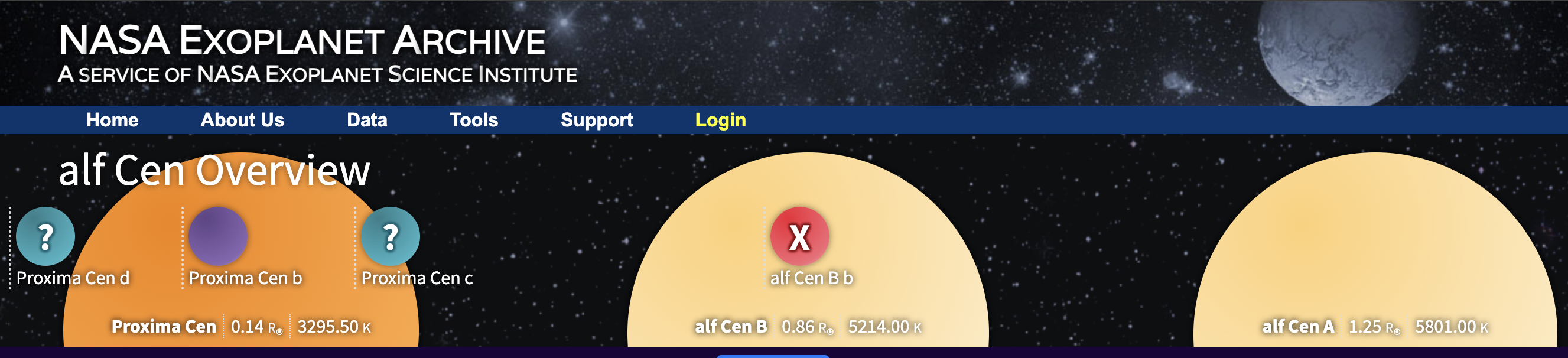}
\caption{A screenshot of the graphical representation of the planetary system architecture at the top of the alpha Centauri system overview page (Sec. \ref{sec:psdata}). This is a triple-star system, with a confirmed planet (shown as a solid purple circle) orbiting proxima Centauri, two candidate planets (shown as blue circles with white question marks) orbiting proxima Centauri, and a false positive planet (shown as a red circle with a white X) orbiting alpha Centauri B.
\label{fig:alphacen}}
\end{figure}

All planetary system data are also found on the individual overview pages for each system. Figure \ref{fig:alphacen} shows a portion of the NEA overview page for the alpha Centauri system.\footnote{alpha Centauri overview page: \url{https://exoplanetarchive.ipac.caltech.edu/overview/alpha\%20cen}} This overview page highlights another aspect of the NEA---if new peer-reviewed analyses are able to identify a false positive scenario for an existing confirmed planet (i.e. that the signal is more correctly attributed to stellar variability or rotation), that object is demoted in the NEA to the disposition `false positive planet', and it no longer appears in the Planetary Systems and Planetary System Composite Data tables. However, the objects remain on the overview pages (c.f. alpha Cen B b in Figure \ref{fig:alphacen}), to accurately represent the evolving understanding of that system. In addition, if new peer-reviewed analyses are unable to recover a previously confirmed planet in the same dataset as it was originally discovered, or recover a signal with a substantially different significance or set of properties in a new dataset, the planet remains confirmed but is flagged as `controversial' until an appropriate false positive scenario is identified or additional data which confirms the planet is obtained. 

\subsubsection{Exoplanet Atmosphere Data}
\label{sec:atmos}
With the launch of JWST in December 2021, the exoplanet community gained unparalleled access to exoplanet atmospheres, both in terms of number of observable planets and the detail with which we might characterize them. Anticipating the increase in both the volume and complexity of exoplanet
atmosphere data, in 2021 the NEA convened a community consultation team with broad atmosphere expertise. This community panel provided a prioritized list of the types of data that would be most useful for the NEA to ingest and serve---the highest priority was a table of high-level, published exoplanet transmission and emission spectra. The new Atmospheric Spectroscopy table\footnote{Atmospheric Spectroscopy table: \url{https://exoplanetarchive.ipac.caltech.edu/cgi-bin/atmospheres/nph-firefly?atmospheres}} combines the previous emission spectroscopy and transmission spectroscopy tables into a single, unified environment that contains all existing atmospheric holdings. The table was expanded earlier in 2024 to include spectra of directly imaged planets. 

\begin{figure}[t!]
\includegraphics[width=\textwidth]{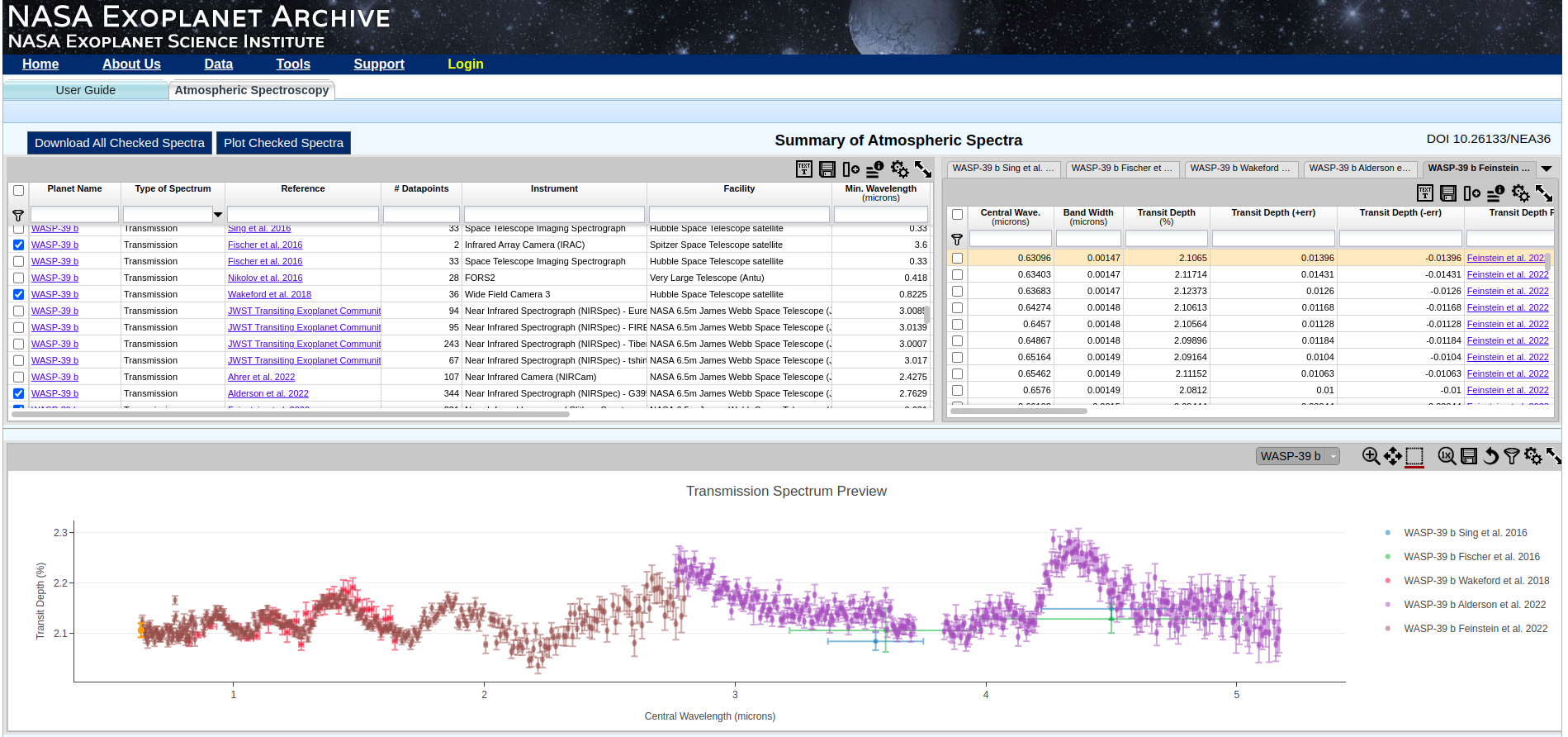}
\caption{An overview of the layout of the NEA Atmospheric Spectroscopy environment (Sec. \ref{sec:atmos}). The table in the top left lists the available spectra in the NEA, the table in the top right lists the data from the selected spectrum in the top left table, and the plot in the bottom panel displays all the selected spectra in the top left table---in this case, Spitzer, Hubble, and JWST transmission spectroscopy of WASP-39 b.
\label{fig:atmos}}
\end{figure}

The data are grouped into discrete spectra and are served in an interactive interface built using the Firefly toolkit developed at Caltech/IPAC \citep[see, e.g.][]{Firefly,joliet201}. The interface allows for the display and interaction between multiple types of data, such as tables, plots, and images (Figure \ref{fig:atmos}), including over-plotting of multiple spectra. The NEA currently serves all published JWST exoplanet spectra; backfill efforts are currently underway to complete our holdings, prioritizing Hubble Space Telescope (HST) and Spitzer data sets. Spectra can be downloaded individually or in bulk in several standard formats. The interface has been designed to allow the NEA to handle the large and varied data products expected from HST, JWST, Pandora (Sec. \ref{sec:nasafuture}), Ariel, and ground-based spectroscopy. It enables the community to find and directly compare spectra across different instruments and for different planets. Future updates to allow users to compare the spectra with opacities and models are planned.

\subsubsection{Specialized Tables}
\label{sec:specdettables}

The comprehensive confirmed planet tables (the Planetary Systems and Planetary System Composite Data tables) contain the super-set of planets detected across ten different techniques. In addition to these tables, we maintain smaller, detection-specific tables, which provide a simplified view into planets observed by a given detection technique, as well as hosting additional parameters specific to that technique (such as Einstein crossing time for microlensing planets, or projected sky separation for directly imaged planets). We currently serve detection-specific tables for microlensing, directly imaged, and transiting planets. We have plans to create radial velocity and astrometry detection tables, the latter to accommodate the influx of expected planets from the ESA Gaia mission. 

Each row of the Planetary Systems table presents a published set of planet parameters and its associated set of stellar parameters. However, the NEA contains useful stellar parameter sets for confirmed planet host stars that are not associated with a published planetary parameter set. Previously, these stellar parameters were only accessible on the individual planetary system overview pages. The NEA recently released a new Stellar Host Parameters table,\footnote{Stellar Host Parameters table: \url{https://exoplanetarchive.ipac.caltech.edu/table/STELLARHOSTS}} serving all NEA stellar parameter sets for stars in systems with confirmed planets, including non-planet hosting companion stars. This table is particularly relevant for the bright nearby stars that have been studied in detail for more than a century, many of which are high-priority targets for the Roman Space Telescope and Habitable Worlds Observatory missions.

\subsubsection{NASA Mission Data Holdings}
\label{sec:nasadata}

The NEA serves as the official archive for many NASA missions and supports additional NASA missions by serving high-level exoplanet discovery and characterization science products.

\begin{itemize}
    \item Kepler/K2 (2009–2018): The NEA, the Kepler project, and the Mikulski Archive for Space Telescopes (MAST) coordinated to group and distribute products produced by the Kepler telescope---MAST hosts the front-end products of the Kepler pipeline (such as the calibrated pixels and de-trended time series) and NEA hosts the back-end products.\footnote{Kepler Mission Products Overview: \url{https://exoplanetarchive.ipac.caltech.edu/docs/KeplerMission.html}} These include the Threshold Crossing Event (TCE) and Kepler Objects of Interest (KOI) lists,\footnote{Kepler Objects of Interest: \url{https://exoplanetarchive.ipac.caltech.edu/table/KOI}} the completeness and reliability products,\footnote{Kepler completeness and reliability products: \url{https://exoplanetarchive.ipac.caltech.edu/docs/Kepler_completeness_reliability.html}} associated false positive tables,\footnote{False Positive tables: \url{https://exoplanetarchive.ipac.caltech.edu/table/KOIFPP},  \url{https://exoplanetarchive.ipac.caltech.edu/table/FPWG}} and multiple simulated data sets.\footnote{Kepler Simulated Data: \url{https://exoplanetarchive.ipac.caltech.edu/docs/KeplerSimulated.html}} In addition, the NEA continues to assign Kepler and K2 names to newly discovered Kepler and K2 exoplanets. The K2 extension to the Kepler mission did not produce back-end products in the same way as Kepler, however, the community independently searched the data for planetary signals, discovering over 1,500 confirmed or candidate planets that are served in the NEA K2 Planets and Candidates table.\footnote{K2 Planets and Candidates: \url{https://exoplanetarchive.ipac.caltech.edu/table/K2PandC}} ExoFOP holds the majority of the Kepler and K2 follow-up data taken by the community, including spectroscopy, high-resolution imaging, derived parameters, and community analyses (Sec. \ref{sec:exofopdata}).
    \item TESS (2018--): The NEA hosts an interactive table of TESS candidates (TESS Objects of Interest,\footnote{TESS Objects of Interest: \url{https://exoplanetarchive.ipac.caltech.edu/table/TOI}}) which is updated each time new literature data are released on the archive. The candidates are integrated throughout the archive, including on system overview pages and from within the Transit and Ephemeris Service.\footnote{Transit \& Ephemeris Service: \url{https://exoplanetarchive.ipac.caltech.edu/cgi-bin/TransitView/nph-visibletbls?dataset=transits}} As with Kepler/K2, ExoFOP has played an integral role in the TESS Follow-up Observation Program (TFOP; Sec. \ref{sec:exofopdata}). The NEA and ExoFOP use the TESS Input Catalog \cite[TICv8.2;][]{Paegert2021}, which incorporates Gaia DR2 as its source catalog alongside Two Micron All-Sky Survey (2MASS) and Wide-field Infrared Survey Explorer (WISE) photometry. The NEA undertook a large effort to cross-match the Kepler, K2, and TESS target stars using the TIC, connecting the Kepler Input Catalog \cite[KIC;][]{Brown2011}, Ecliptic Plane Input Catalog \cite[EPIC;][]{Huber2016} and TIC identifiers for hundreds of thousands of objects, allowing ExoFOP and the NEA to easily share data across services.
    \item JWST (2021--): The NEA supports the exoplanet community by ingesting and serving published, high-level JWST science products. This currently includes planet discoveries, updated planet and system parameters, and exoplanet transmission, eclipse, and direct emission spectroscopy. We make the rapid ingestion and release of JWST data one of our highest priorities for our regular data releases, only superseded by the release of newly confirmed planets. The complex and inventive use of JWST by the exoplanet community continues to grow and as the community publishes them, the NEA will also include exoplanet phase curves and detections of atomic and molecular species (Sec. \ref{sec:future}).
    \item ASTERIA (2017--2019): The Arcsecond Space Telescope Enabling Research In Astrophysics (ASTERIA) mission was a technology demonstration CubeSat led out of the Massachusetts Institute of Technology (MIT) and NASA's Jet Propulsion Laboratory (JPL) to perform high-precision photometry of a small number of exoplanet systems. Processed light curves and images for the two planet-hosting stars, 55 Cnc and HD 219134, are archived at the NEA.\footnote{ASTERIA Mission: \url{https://exoplanetarchive.ipac.caltech.edu/docs/ASTERIAMission.html}}
    \item Spitzer (2003--2020): The NEA is currently serving nearly 170 transmission and emission spectra from Spitzer, as well as confirmed planets and updated planet parameters. We also serve the Spitzer Kepler Survey \cite[SpiKeS;][]{Werner2021} high-precision photometry of the $\sim$200,000 stars monitored by Kepler.
    \item NASA Keck time: The NEA contains data for 189 planets discovered using W. M. Keck Observatory data, and users have uploaded $\sim$1,200 images and $\sim$5,500 spectra to ExoFOP from Keck instruments obtained in follow-up of NASA mission candidates from Kepler, K2, and TESS.
    \item CUTE (2021--): The Colorado Ultraviolet Transit Experiment (CUTE) mission is a CubeSat led out of the University of Colorado to perform low-resolution, near-ultraviolet spectroscopy of a small number of transiting planets. The NEA hosts the high-level science products from CUTE, including processed light curves and transmission spectra, as they become available. The first release of CUTE data was in 2024.
    \item Habitable Worlds Observatory: Habitable Worlds Observatory (HWO) is a future NASA flagship mission to directly image and obtain spectra of Earth-like planets in the habitable zones of Sun-like stars. Astro2020 recommended that a science and technology maturation program begin studying the concept for HWO, and NASA has since established a HWO project office. To prepare for HWO, the NEA hosts several potential target lists.\footnote{HWO Target Lists: \url{https://exoplanetarchive.ipac.caltech.edu/docs/MissionStellar.html}} These include the HWO ExEP Precursor Science Stars List, produced by the Exoplanet Exploration Program (ExEP) Office and other lists assembled by the community. The NEA is also planning to host precursor science products as they become available.
\end{itemize}

\subsubsection{Contributed Data}
\label{sec:contributed}

Part of the NEA's support for the exoplanet community is to provide a platform for a growing number and variety of community-contributed data sets. Table \ref{tab:contributed} summarizes the contributed data sets hosted by the NEA. Here we highlight several datasets. 

Critical to understanding planets is understanding the stars that host those planets. Stellar abundances are correlated with planetary formation and evolutionary processes \citep{Schulze2021}, and the community has done substantial work to understand the elemental composition of planet-hosting stars. The NEA has partnered with the Hypatia Catalog\footnote{Hypatia Catalog: \url{https://www.hypatiacatalog.com/}} to integrate stellar chemical abundances into our planetary system overview pages. The NEA and Hypatia collaborated to develop an efficient method for automated retrieval and presentation of the latest data from the Hypatia Catalog.

Finally, the NEA hosts over 130,000,000 photometric time series, summarized in Table \ref{tab:timeseries}. These include both space-based (Kepler and CoRoT) and ground-based (Super Wide-Angle Survey for Planets [SuperWASP], Kilodegree Extremely Little Telescope [KELT], XO (not an acronym, as in ``exoplanet''), and the Hungarian-made Automated Telescope Network [HATNet]) transit surveys. There are also two large microlensing surveys of the Galactic bulge, United Kingdom InfraRed Telescope (UKIRT), and Microlensing Observations in Astrophysics (MOA), in support of NASA's Nancy Grace Roman Space Telescope microlensing survey (Sec. \ref{sec:nasafuture}). Each time series dataset features a metadata file for the full dataset and individual files for each light curve. The metadata file contains a summary of the data set as a whole, as well as a summary of each individual light curve. Through the archive infrastructure, users may query the metadata file to search the data set by metadata. The light curves can be viewed in the NEA's Time Series Viewer or uploaded into the periodogram and \texttt{EXOFAST} fitting services (Sec. \ref{sec:fitting}).

\begin{table}
\centering
\begin{tabular}{|l|l|l|}
\hline
Data set & Type of Data & Volume \\
\hline \hline
ExEP HWO Precursor & Curated short list of stars likely to be targeted by HWO & \\ 
Science Stars & with associated parameters & 164 stars\\ \hline
HWO Preliminary Input & Curated long list of stars potentially targeted by HWO with & \\
Catalog (HPIC) & associated parameters & $\sim$13,000 stars\\ \hline
FDL/INARA &  Simulated exoplanet atmosphere spectra & $\sim$3,000,000 spectra \\ \hline
FDL/PyATMOS &  Simulated Earth-like exoplanet atmosphere spectra & $\sim$124,000 spectra \\ \hline
MARGE-HOMER & Simulated hot-Jupiter atmosphere spectra & $\sim$3,500,000 spectra \\ \hline
 &  & 10 abundances/star, \\
Hypatia & Stellar abundances for 10 high-priority elements & 1600 stars \\ \hline
 & A volume-limited sample of stars and brown dwarfs within 20pc & \\
20pc Brown Dwarf Census & of the Sun & $\sim$4,400 objects \\ \hline
PEPSI Exoplanet Transit & High-resolution transmission and emission spectroscopy of hot & \\ 
Survey & Jupiters & $\sim$90 spectra \\ 
\hline
\end{tabular}
\caption{Community-contributed data sets in the NASA Exoplanet Archive}
\label{tab:contributed}
\end{table}

\begin{table}
\centering
\begin{tabular}{|l|l|}
\hline
Instrument & Volume \\
 \hline \hline
United Kingdom InfraRed Telescope (UKIRT) & $\sim$100,000,000 time series\\
Super Wide Angle Search for Planets (SuperWASP) & $\sim$18,000,000 time series\\
Microlensing Observations in Astrophysics (MOA) Survey &  $\sim$2,400,000 time series \\
Kilodegree Extremely Little Telescope (KELT) & $\sim$7,000,000 time series\\
CoRoT asteroseismology \& exoplanet surveys & $\sim$160,000 time series\\
Cluster survey data & $\sim$140,000 time series\\
XO Survey & $\sim$2,000 time series\\
Hungarian-made Automated Telescope Network (HATNet) & $\sim$6,000 time series\\
Kepler & $\sim$2,895,000 time series\\
 \hline
\end{tabular}
\caption{Time series data sets in the NASA Exoplanet Archive}
\label{tab:timeseries}
\end{table}

\subsection{ExoFOP Data Holdings}
\label{sec:exofopdata}

The Exoplanet Follow-up Observing Program (ExoFOP) website is a complementary service to the NEA. Whereas the goal of the NEA is to archive and serve highly curated data and derived data products from the published literature and NASA missions, the goal of ExoFOP is to
provide an open access sandbox in which the community can share, collaborate, and coordinate. By enabling observers to prioritize and plan the efficient and effective allocation of valuable follow-up resources, ExoFOP supports the characterization of targets (stars and planets) and the confirmation of NASA mission candidates.

\subsubsection{NASA Mission Support}

ExoFOP began life in service to the Kepler mission as the Kepler Follow-up Observing Program (KFOP) website. Initially, access was restricted to the Kepler science team for their coordination and follow-up of the (then private) KOIs. With the public release of KOIs, the community became heavily involved in and began leading the follow-up of Kepler candidates, and with the evolution of Kepler into the K2 mission, the wider community began identifying and following up K2 planet candidates. KFOP evolved to become a fully public service---the Community Follow-up Observing Program (CFOP)---with all data visible to all visitors to the website. This led to a significant increase in users, activity, and citations to the service. KFOP and CFOP were essential services to the Kepler and K2 missions, and there are still nearly 3,000 planet candidates from these missions that are as yet unconfirmed. Active follow-up and analysis of these candidates continues---70 new planets were confirmed in Kepler data in 2023 alone. There are ongoing efforts to synthesize data from the three missions to refine the parameters of known planets.

Building on the success of Kepler and K2, ExoFOP-TESS was launched to support the NASA TESS mission as an explicitly defined, publicly accessible, archive for the TESS Follow-Up Observation Program (TFOP). Achieving the primary mission science requirement of measuring the masses of 50 planets smaller than Neptune, and extended mission goals of continuing to expand the sample of bright, well-understood exoplanets for JWST characterization, requires substantial, sustained follow-up coordination and collaboration across many teams, instruments, and techniques. The TESS community uploads time-series photometry, spectroscopy, radial velocity, and high-resolution imaging data as part of the candidate confirmation and characterization process (Fig.~\ref{fig:exofop}). The openness of ExoFOP and the ability of the NEA to put the TESS candidates in context with comprehensive literature values and data, alongside powerful fitting and planning tools, has enabled significant participation and leadership in TESS science by early career researchers. 

\begin{figure}[t!]
\centering
\includegraphics[width=0.8\textwidth]{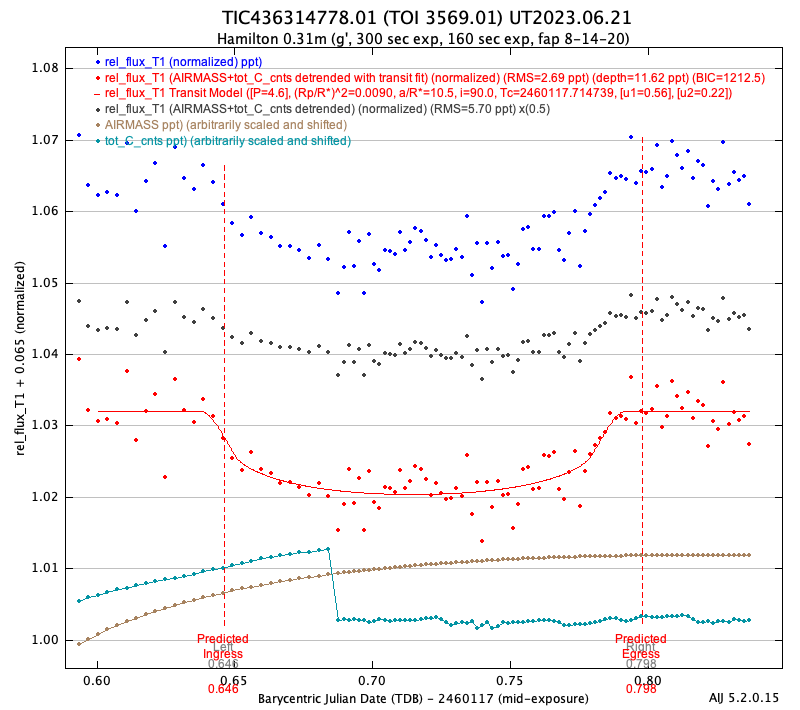}
\caption{Example of ground-based time-series photometry of TOI-3569 obtained by the TESS Follow-Up Observation Program and shared with the community on ExoFOP (Sec. \ref{sec:exofopdata}).
\label{fig:exofop}}
\end{figure}

\begin{figure}[t!]
\centering
\includegraphics[width=\textwidth]{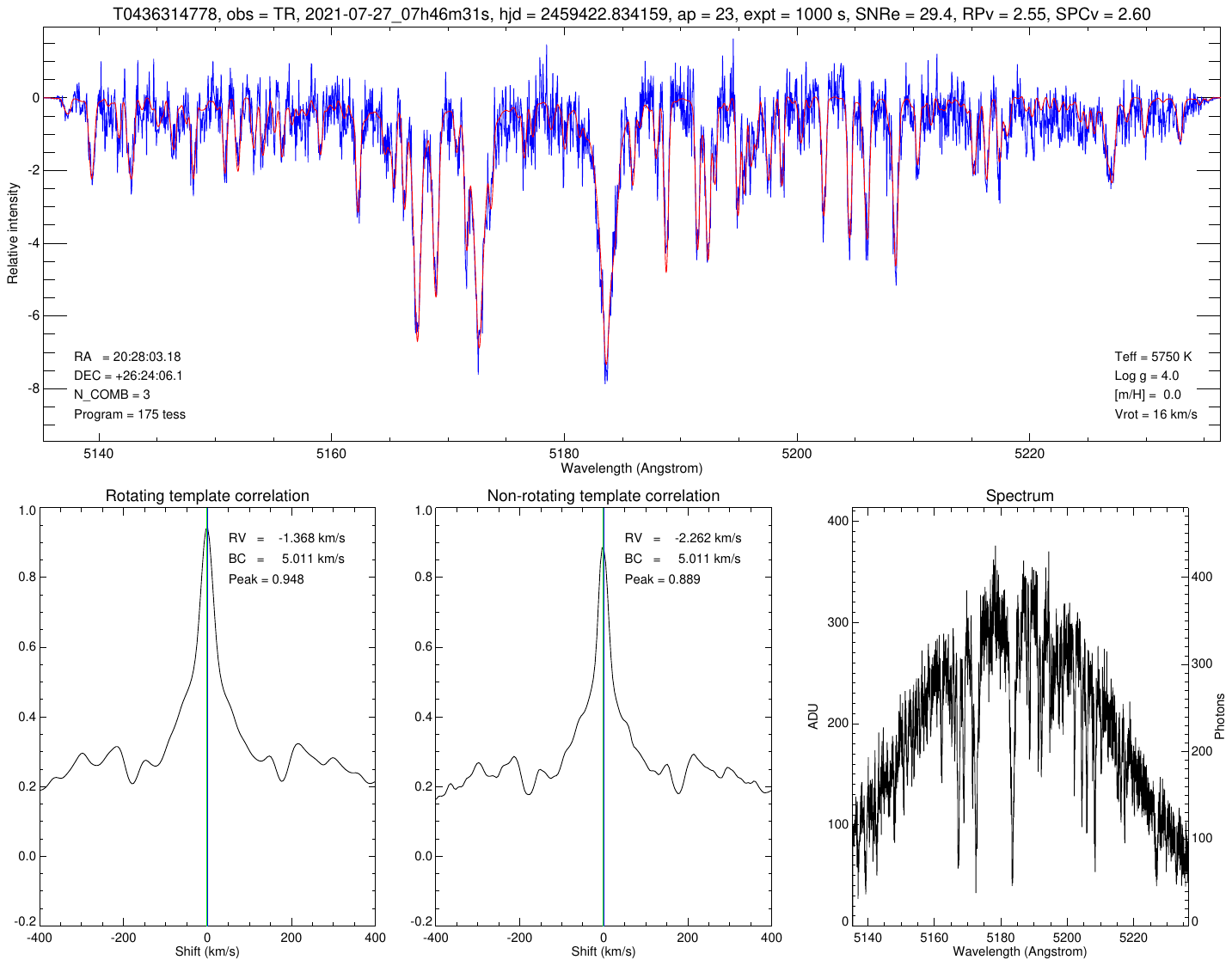}
\includegraphics[width=0.5\textwidth]{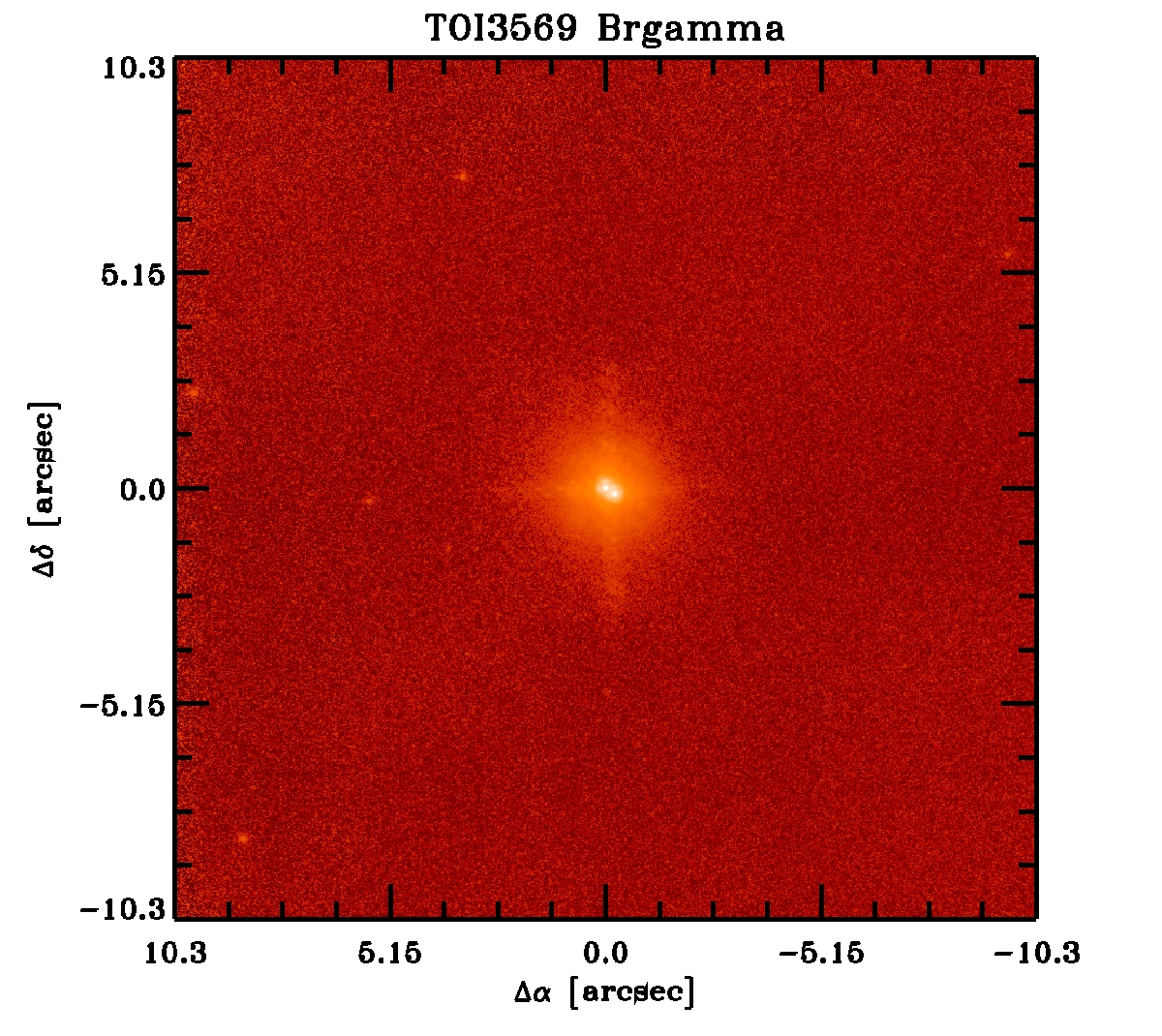}
\caption{Examples of reconnaissance spectra (left), and high-resolution imaging of TOI-3569 (right) obtained by the TESS Follow-Up Observation Program and shared with the community on ExoFOP (Sec. \ref{sec:exofopdata}).
\label{fig:exofop}}
\end{figure}

In 2022, KFOP, CFOP, and ExoFOP-TESS were merged into a single ExoFOP portal. All existing content, including user profiles and upload histories, was preserved. ExoFOP has over 1,600 registered users (registration is required to upload but not to view publicly available data), and nearly one million files have been uploaded across the Kepler, K2, and TESS follow-up observing programs. Over 70,000 observing summaries have been uploaded, with $\sim$30,000 associated notes.

In addition to the TESS project’s TOIs, community members can search the publicly available TESS data at MAST for their own signals. ExoFOP provides a mechanism for users to initiate Community TESS Objects of Interest (CTOIs), with a required minimal set of parameters that allows the TESS Science Office to search for the signal and, if appropriate, promote the signal to a TOI. At present, the community has identified more than 3,400 CTOIs with 20\% of them having been converted to TOIs by the TESS Project. Some ExoFOP functionality is currently limited to the $\sim$20,000 candidates across Kepler, K2, and TESS; in the future we will allow users to initiate and interact with their own independently discovered candidates and project-specific target lists (see Sec. \ref{sec:future}).

Finally, ExoFOP has been adapted to support NASA's contribution to the Ariel mission. ExoFOP serves potential Ariel targets---known confirmed exoplanets and TESS candidates that are likely to be Ariel targets---enabling sharing of data and results of observational work necessary to characterize the stars and planets as part of the Ariel Preparatory Science Program. The Ariel Mission and Science Consortium is undertaking a preparatory science program to support target prioritization and NASA is enabling US community participation in this program as part of the NASA-ESA partnership in Ariel. ExoFOP---as it has done with Kepler, K2, and TESS---is serving as the target and data collection point for the contribution.

We have prioritized making connections, exploiting synergies, and ensuring uniformity between the data holdings at the NEA and ExoFOP. ExoFOP hosts the full TIC table, and the system parameters (positions, parallaxes and distances, proper motions, and photometry) for objects in the NEA are drawn from that resource. ExoFOP queries the TESS Science Office twice daily for new and updated TOIs, and the NEA automatically updates its TOI table from the ExoFOP holdings at each literature data release. The NEA system overview pages link to the corresponding overview page at ExoFOP, and the ExoFOP overview pages show the confirmed planet information from NEA. In addition to linkages with the NEA, ExoFOP pages link to the InfraRed Science Archive (IRSA) finding charts, the Set of	Identifications, Measurements and Bibliography for Astronomical Data (SIMBAD), MAST, and Keck Observatory Archive (KOA) databases, and several external tools. All ExoFOP users are expected to follow the ExoFOP Data Use and Professional Conduct Policy,\footnote{ExoFOP Data Use and Professional Conduct Policy: \url{https://exofop.ipac.caltech.edu/tess/pcp.php}} which provides direction on acknowledgment, citation and/or co-authorship for use of data uploaded to ExoFOP.

\subsection{Data Access} \label{sec:access}

The data served by the NEA and ExoFOP is not useful if it is not readily discoverable, accessible, well-sourced and documented, and available in commonly used formats. As the community continues to move toward more program-driven ways to access data, and science platform infrastructure begins to take shape, it is crucial that data archives adopt standard ways to access data. The NEA endeavors to make data holdings available in multiple ways, to suit users' experience and workflows. In addition to the web interface (see, e.g., access functionality described in Sec. \ref{sec:icetables}), we offer the following ways to access our holdings:


\begin{enumerate}
\item \emph{TAP}: Programmatic access to all interactive  tables\footnote{Using TAP: \url{https://exoplanetarchive.ipac.caltech.edu/docs/TAP/usingTAP.html}} is provided via an Application Programming Interface (API) using International Virtual Observatory Alliance (IVOA)-compliant Table Access Protocol (TAP) standards.  Adoption of TAP across the board enables broad interoperability with the NEA, and allows users to take advantage of efforts in the wider community to build tools and services that use TAP-provided data, including the other astrophysics archives. The NEA's TAP service uses IVOA Astronomical Data Query Language (ADQL), allowing \texttt{select} and \texttt{where} clauses, and the specification of the output format. Here is an example query to return the names, masses, and positions of confirmed Earth-sized planets with masses 0.5--2~$M_{\oplus}$ in comma-separated value format:
\indent \texttt{https://exoplanetarchive.ipac.caltech.edu/TAP/sync?query=select+pl\_name,pl\_masse,ra,dec+from+
ps+where+upper(soltype)+like+'\%CONF\%'+and+pl\_masse+between+0.5+and+2.0\&format=csv}

\item \emph{astroquery}: \texttt{Astroquery} \citep{Astroquery} is a coordinated set of tools within \texttt{astropy} for querying astronomical databases, and includes a NASA Exoplanet Archive module.\footnote{Astroquery: \url{https://astroquery.readthedocs.io/en/latest/ipac/nexsci/nasa_exoplanet_archive.html}} The module uses NEA's TAP service described above to access tabular data from within a Python kernel. 

\item \emph{Bulk download}: For large datasets that are not efficiently served in an interactive web interface, we provide scripts for downloading data products in bulk,\footnote{Bulk Data Download: \url{https://exoplanetarchive.ipac.caltech.edu/bulk_data_download/}} which requires installing the GNU {\bf \texttt{wget}} software package.
\end{enumerate}

\section{NASA Exoplanet Archive Tools} \label{sec:tools}

One goal of the significant recent refactoring of the NEA underlying architecture was to allow the NEA to more easily deploy new views into the data, such as the Planetary Systems Composite Data table, and to develop new services that take advantage of the data, such as the Atmosphere Spectroscopy table (Sec. \ref{sec:atmos}).

\subsection{Interactive tables}
\label{sec:icetables}

The NEA has unified all existing tables under a single type of interactive interface, simplifying usage and look and feel across the archive. These interactive tables offer a plethora of functionality, including:
\begin{itemize}
\item Column selection: Many interactive tables only show a portion of the available columns. Additional columns can be selected by clicking the `Select Columns' widget in the top left of the table. Columns can be added, removed, and reordered within the interface.
\item Row selection: Rows can be filtered and sorted, as well as individually selected and deselected with check boxes. All row selections can be reset with the widgets in the bottom right of the table. 
\item Download: Tabular data can be downloaded either in its entirety or in its filtered state in a number of formats through the `Download Table' widget at the top left of the table.
\item Visualization: Tabular data can be sent to our plotting visualization tool either in its entirety or in its filtered state via the `Plot Table' widget in the top left of the table (see Sec. \ref{sec:plotting} for details of that tool).
\item Documentation: Pages describing the available columns, how to use the interactive tables, and any dataset-specific information are available in the `View Documentation' widget in the top left of the table.
\end{itemize}

\subsection{Visualization}
\label{sec:plotting}

\begin{figure}[t!]
\includegraphics[width=\textwidth]{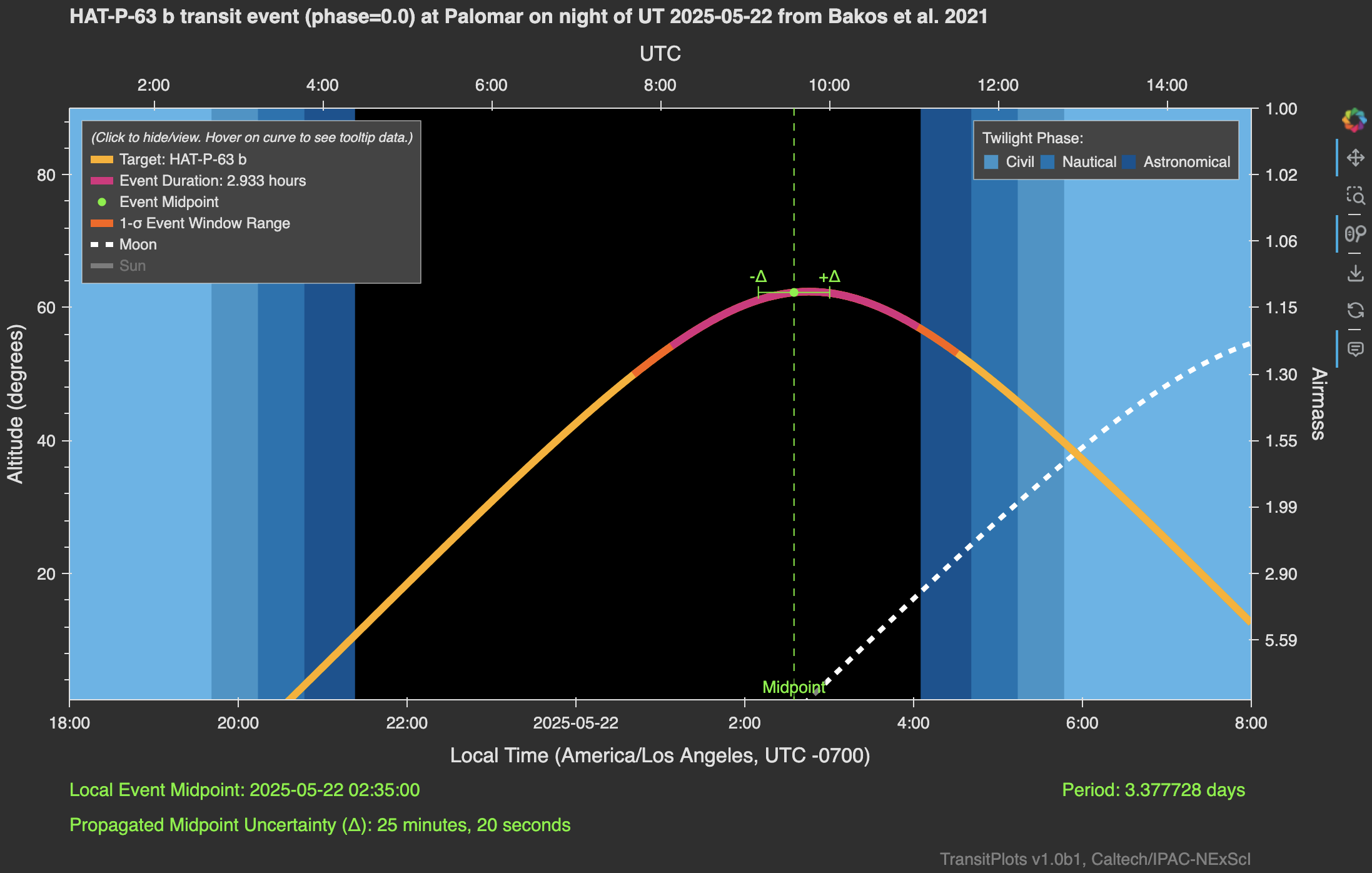}
\caption{An example airmass plot provided with the results of the Transit and Ephemeris Service (Sec. \ref{sec:transitservice}). The plot shows the altitude/airmass of an object as it rises and sets over the course of a night, with any predicted transits highlighted in pink, and the 1$\sigma$-uncertainty on the predicted transit midpoint time shown in green. This example shows the visibility of the transit of HAT-P-63~b as seen from Palomar Observatory on the night of 2025-05-22.
\label{fig:airmass}}
\end{figure}

The NEA has recently made major improvements to plotting and visualization services. Contributed data sets (see Sec. \ref{sec:contributed}) are released using IPAC’s Firefly toolkit, which allows real-time interaction between multiple data types (e.g., tables and plots) in one interface. In response to community feedback, we have moved the development of visualization services used in our main data tables (e.g., Planetary Systems) to Python, using {\bf \texttt{Plotly} and \texttt{Dash}}. Recently, the NEA released the first versions of interactive scatter and histogram plots for our major interactive tables. In an improvement over our previous plotting tool, users can visualize additional variables mapped to symbol sizes and colors, select from preset color palettes and presentation modes, hover over individual data points for additional information, and click on data points to open the object overview page. In addition, the NEA released an Airmass Visualization tool, built using \texttt{Bokeh}, within the Transit and Ephemeris Service (Sec. \ref{sec:transitservice}) to help users plan observations (Figure \ref{fig:airmass}). For a given night, the tool shows altitude and air-mass values for the specified target, with twilight indicators and overlays for the Sun, Moon, and orbital events (such as transits) that can be toggled on and off. It has been extremely effective for the extensive TFOP follow-up efforts to achieve the primary TESS science requirements.

\begin{figure}[t!]
\includegraphics[width=\textwidth]{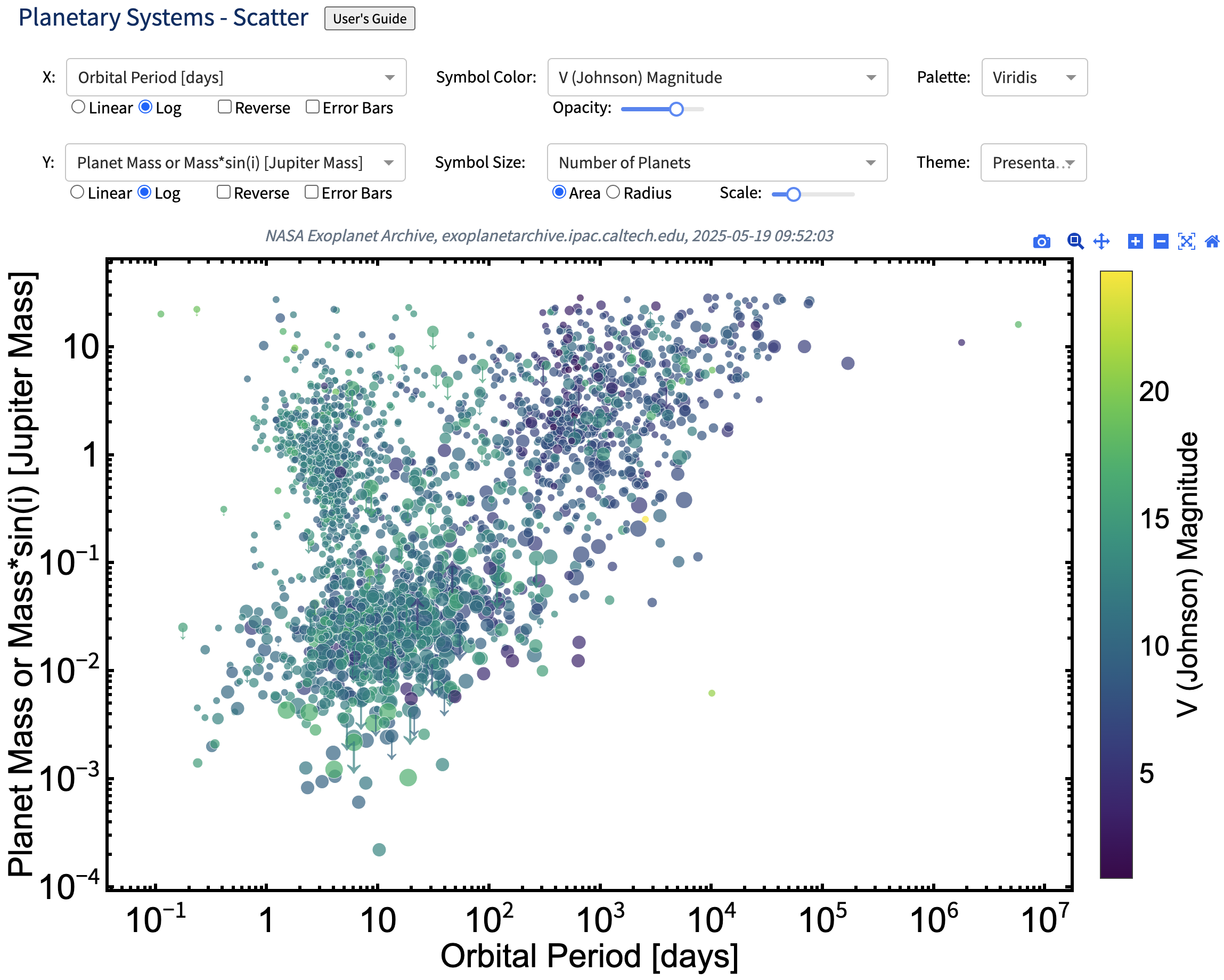}
\caption{A screenshot of the interactive table plotting tool, which is accessible from the top menu bar of each interactive table (Sec. \ref{sec:icetables}). Users can select $x$ and $y$ values from a subset of the (typically numerical) columns in the originating table, and also select additional columns to control the size and color of the plotted values. The plots are available in a range of color palettes and a set of themes (Light, Dark, and Presentation, shown here). Additional functionality includes zooming, panning, and downloading the final plot. Individual points can be hovered over for a brief summary of the plotted data, and clicked to be taken straight to the overview page for that object.
\label{fig:newplots}}
\end{figure}

\subsection{Transit and Ephemeris Service}
\label{sec:transitservice}

The Transit and Ephemeris Service\footnote{Transit \& Ephemeris Service: \url{https://exoplanetarchive.ipac.caltech.edu/cgi-bin/TransitView/nph-visibletbls?dataset=transits}} is an observation planning tool that calculates the times and observability of orbital events (such as transits, eclipses, and quadratures) for planet and planet candidate orbital solutions in the NEA, or custom user-defined orbital solutions. Ephemeris uncertainties are propagated forwards to the desired observing time to provide an estimated window for the orbital event. The locations of many ground-based observatories are pre-loaded in the tool, as is JWST in its L2 orbit, and users can also enter custom locations. This service is one of the NEA’s most used applications, with tens of thousands of queries per year, and significant increases in usage before major proposal deadlines (such as JWST). The tool was recently modified to operate in a background mode to handle the ever-growing number of confirmed planet and planet candidates, and to be able to be called programmatically as part of the future development (Sec. \ref{sec:future}).

\subsection{Fitting tools}
\label{sec:fitting}

The NEA provides several interfaces for fitting time series data, including photometry and radial velocities. The first is an interactive interface to the \texttt{EXOFAST} fitting code \citep{Eastman2013}. The NEA's version of \texttt{EXOFAST}\footnote{\texttt{EXOFAST}: \url{https://exoplanetarchive.ipac.caltech.edu/cgi-bin/ExoFAST/nph-exofast}} offers the same IDL-based calculations as the original code, and also provides sufficient back-end computing resources to enable Markov Chain Monte Carlo (MCMC) analysis. The interface allows users to use up-to-date NEA parameters as inputs and priors to the fits, and allows users to provide both photometric and radial velocity time series files. As output, the service provides plots of the best-fit models overlaid on the time series files, plots of the covariance matrices and posterior probability distributions, and a LaTeX-formatted table of fitted parameters and uncertainties.

In addition to \texttt{EXOFAST}, the NEA provides a simpler periodogram tool.\footnote{Periodogram tool: \url{https://exoplanetarchive.ipac.caltech.edu/cgi-bin/Pgram/nph-pgram}} The tool allows users to upload time series data, specify the period range to be searched, and select from three periodogram algorithms for finding periodic signals in the data.

\section{User Support}
\label{sec:user}

The NEA and ExoFOP consistently engage the broader community to identify the most useful data to be serving and develop the tools needed for use with those data. The NEA User Panel is assembled from experts across a
variety of exoplanet sub-fields, detection techniques, and institutions.\footnote{NASA Exoplanet Archive User Panel: \url{https://exoplanetarchive.ipac.caltech.edu/docs/users_group.html}} The panel meets annually to hear updates on
progress, to help prioritize short-term goals, and to provide guidance on longer range plans. In addition to these formal consultations, NEA and ExoFOP staff attend the summer and winter American Astronomical Society (AAS) meetings and annual Division of Planetary Sciences (DPS), Astronomical Data Analysis Software \& Systems (ADASS), and IVOA meetings to proactively interface with attendees. This allows us to demonstrate new features at the NEA and ExoFOP, ask questions about user experiences, and solicit constructive feedback on user interface and design. In addition, the NEA includes a web-based, ticketed help desk system where users can ask questions, enter bug reports, and provide requests to include contributed data, to add functionality, or to clarify data inaccuracies.

To communicate updates, the NEA produces a newsletter of news and announcements that is emailed at each data release. Announcements are also made through the NASA Exoplanet Science Institute (NExScI) and ExEP email lists. The NEA also shares a summary of the previous month’s news items in the ExoPlanet News,\footnote{ExoPlanet News: \url{https://nccr-planets.ch/exoplanetnews/}} a monthly electronic newsletter produced by the European exoplanet community. Finally, the NEA has a social media presence on X (formerly Twitter) and Facebook, for engaging directly with professional and citizen scientists who use the NEA's data and services. The NEA accounts have a distinct use case and audience compared to the broader and more general exoplanet social media accounts for the general public served by NASA's Exoplanet Exploration Program.

\section{Future Plans} \label{sec:future}
\subsection{Literature data}
\label{sec:literature}

The next decade will mark an unprecedented increase in exoplanet discovery and characterization. The community’s focus will continue bifurcating into deep characterization of the closest systems, and the study of populations of planets and how the populations are sculpted by stellar and system properties. The NEA and ExoFOP will continue to support these analyses as the data increase in scale and complexity. Table \ref{tab:yields} summarizes the expected yields of current and planned exoplanet surveys, with over 200,000 planets predicted to be found, which can also be seen in Figure \ref{fig:counts}.

In addition to the machine-learning paper classifier that the NEA deploys to scan the published literature for new data each day (Sec. \ref{sec:psdata}; Susemiehl in prep.), the NEA is developing machine learning tools to extract data from the papers themselves. These tasks include identification of the objects in the paper, detection and discovery methods and facilities, and extraction, formatting, and validation of stellar and planet parameters provided in tables. The NEA is also preparing formatted text file templates for authors to use when submitting their papers, which can be read and ingested by NEA software. Use of these templates would both accelerate the process for including papers in the NEA, and improve the accuracy of identification and transcription of values from papers into the database tables.

\begin{table}
\begin{tabular}{|l|l|l|}
\hline
Survey & Predicted Yield & References \\
 \hline \hline
NASA TESS mission & 12,000--15,000 & \cite{Barclay2018,Kunimoto2022}\\
ESA Gaia mission & $\sim$70,000 & \cite{Perryman2014}\\
Roman Space Telescope Microlensing Survey & $\sim$1,400 & \cite{Penny2019}\\
Roman Space Telescope Transit Survey & 60,000--200,000 & \cite{Montet2017,Wilson2023}\\
ESA PLATO mission & $\sim$13,000 & PLATO Definition Study Report*\\
China National Space Agency Earth 2.0 mission & $\sim$30,000 & \cite{Ge2024}\\
\hline
\end{tabular}
\caption{Predicted yields of current and planned exoplanet surveys. *\url{https://platomission.com/wp-content/uploads/2018/05/plato2-rb.pdf}}
\label{tab:yields}
\end{table}

In addition to detection surveys, the community is firmly in the era of increasingly detailed exoplanet characterization, from atmospheres to surfaces and even to interiors. One goal of the NEA and ExoFOP is to capture and serve value-added ancillary exoplanet system data beyond the basic discovery information through specialized detection tables (Sec. \ref{sec:specdettables}), the new atmosphere environment (Sec. \ref{sec:atmos}), and our link to Hypatia stellar abundances. The NEA will continue to serve published, high-level science products from JWST and Hubble, and toward the end of the decade, from the ESA Ariel mission, in which NASA is a partner through the Contribution to Ariel Spectroscopy of Exoplanets (CASE) project. Ariel will survey $\sim$1,000 exoplanet atmospheres to uncover broad trends in atmosphere composition, structure, and evolution, and to enable population studies of different atmosphere types. In addition to transmission, eclipse, and direct imaging spectroscopy, the NEA has plans to support spectroscopic phase curves, a table of atoms and molecules detected in exoplanet atmospheres, and a table of high-resolution spectroscopy meta-data.

\subsection{Upcoming NASA Mission Support}
\label{sec:nasafuture}

The NEA will continue to support the TESS mission through its extended missions, as well as a number of upcoming NASA missions:

\begin{itemize}
    \item Pandora (2025--2026): The NASA Pioneers mission, Pandora, is a SmallSat scheduled to launch in 2025. The primary science goal of Pandora is investigating and understanding the transit light source effect, whereby heterogeneities on the stellar surface produce features in planet transmission spectroscopy that can be mistaken for features in the planet’s atmosphere \citep[see, e.g., the ambiguity in the source of water in the JWST transmission spectrum of GJ 486 b,][]{Moran2023}. NExScI is the Data Archive Center for Pandora, and the NEA will serve the raw and processed science data products from the mission to the community and provide a permanent archive for these products. Analogous to the Kepler mission products, the data will be served both together on Pandora-specific pages and integrated throughout the rest of the archive holdings where appropriate.
    \item Nancy Grace Roman Space Telescope (2026--): In support of Roman, the NEA is hosting microlensing contributed data sets (see Sec. \ref{sec:contributed}). In addition, the NEA is supporting two funded NASA ROSES Roman Research and Support Participation Opportunities---the microlensing Project Infrastructure Team (PI Gaudi) for which the NEA will host the high-level science products produced as part of the three planned data challenges, and the transiting Wide-Field Survey team (PI Quintana), for which the NEA will host the simulated source catalog, photometric time series, and detection diagnostic plots. Finally, NEA plans to update the implementation of \texttt{EXOFAST} (Sec. \ref{sec:tools}) to \texttt{EXOFAST III}, which will include Mulens-Model \citep{Poleski2019}, a microlensing model-fitting code. The update will include the capability of calling and fitting the microlensing time series that will be produced and served by the IPAC Roman Science Support Center (SSC). The NEA plans to communicate closely with the Roman project regarding the format and timeline of the delivery of planets and planet candidates for efficient ingestion into the NEA.
    \item Landolt (2029-2030): The NASA Pioneers mission, Landolt, is a smallsat scheduled for launch in 2029. The primary science goal of Landolt is to utilize flux calibrated, downlooking beacons to provide a multi-wavelength absolute flux calibration source for ground-based observing. The absolute flux calibration will enable improvements of stellar parameters to 1--2\%, and thus make measurement error the dominant source of uncertainty in the determination of planetary parameters. The ground-based observations and the derivation of accurate and precise stellar parameters and planetary parameters will be archived at the NEA and ExoFOP.
    \item Habitable Worlds Observatory (2040s--): Support for HWO currently includes the following: (i) hosting of potential target lists for HWO published by ExEP and by the community (Sec. \ref{sec:nasadata}); (ii) development of the Target List Management capability at ExoFOP to support characterization of potential HWO targets (§4.2.1.6); and (iii) identification and hosting of precursor science products produced by the community (§4.1.2.2).
    
\end{itemize}

\subsection{ExoFOP Future Plans}

ExoFOP will expand its capabilities to allow for more broad, generic support of the community’s exoplanet characterization efforts beyond Kepler, K2, and TESS. This includes: (i) opening up the ability for users to upload observation summaries, notes, files, and parameters on planets and planet candidates that are not necessarily TOIs, CTOIs, or KOIs; (ii) providing more linkages to existing content at the NEA and external archives, such as MAST; (iii) providing standardized API access to the data available at ExoFOP for ease of discovery, and integration in user’s own ecosystems; and (iv) providing a new target list management system. As the Gaia mission releases astrometric planetary candidates and the PLATO mission makes their transiting planet candidates public, ExoFOP will ingest these candidates and their associated properties to enable organized and efficient community follow-up. 

In response to the Exoplanet Program Analysis Group (ExoPAG) Science Analysis Group (SAG) \#22 report,\footnote{ExoPAG SAG \#22 report: \url{https://arxiv.org/pdf/2112.04517}} which found the assembly, curation, and presentation of stellar parameters for stars that would be likely targets of future missions (such as HWO) would be beneficial to those missions, ExoFOP is developing a new target list management service. This service is an evolution of the existing My Targets service, which allows users to track updates to a previously uploaded list of TIC IDs. The Target List Management service will allow teams to build custom target lists of stars (or planets), to select parameters of interest for those targets, to identify and track gaps in those parameters (such as needed observations or derived parameters), and to coordinate and collaborate on deriving as robust an understanding of the potential targets as possible. As part of this effort, ExoFOP will be updated to include those stellar parameters identified as important in the SAG 22 report that are not currently supported by ExoFOP, including abundances, additional magnitudes, disk information, and stellar companion information.

\subsection{Data and Tool Access}

The NEA plans to develop Python-wrapped APIs around the tool services to enable users to utilize the NEA tools from their own environments alongside NEA data holdings. The tools, while called from the client, will run locally on the NEA server, enabling quick access to the data and access to the processing power of the data center environment or cloud. The tool APIs will take advantage of the TAP service to access data in the NEA. Integrating the NEA’s data with publicly available fitting codes, such as \texttt{EXOFAST}, will serve as pilot projects that will build out the required infrastructure. Once in place, the NEA tools, starting with the Transit and Ephemeris Service, will be modified to support API calls. This is a critical step toward a longer-term plan of enabling the use of NEA and ExoFOP data and services in the NASA implementation of the Fornax science platform, the development of which is being led by the High Energy Astrophysics Science ARChive (HEASARC), IRSA, and MAST.

The development for the interactive plotting (Sec. \ref{sec:plotting}) and the data and tool APIs will provide the NEA with the building blocks to develop Python-oriented libraries and science-oriented tutorial notebooks, to extend the notebooks already provided by ExoFOP. These notebooks could be utilized by the community to access, explore, and visualize the data holdings of the NEA and ExoFOP. Notebooks will help the community with specific workflows, but also enable them to better understand the content and its context---whether published parameters, data products or community data on the NEA or ExoFOP. Publicizing NEA and ExoFOP data through existing tools within the astronomical Python community, such as PyVO, will significantly improve the findability of our extensive data holdings.

\section{Acknowledging the NEA and ExoFOP}
\label{sec:acks}

The NEA and ExoFOP are freely available services for the exoplanet community. However, it is useful for future prioritization of data and tools to track their usage by the community, so both services request that citations or acknowledgments be included in publications which make use of the services. For the NEA, this paper can be cited, or the following language can be included in the paper acknowledgments: 

\begin{displayquote}
``This research has made use of the NASA Exoplanet Archive, which is operated by the California Institute of Technology, under contract with the National Aeronautics and Space Administration under the Exoplanet Exploration Program.''
\end{displayquote}

Individual sections of the NEA also have assigned Digital Object Identifiers (DOIs) which can be referenced to cite a specific data source; for example, the Planetary Systems table has the specific DOI 10.26133/NEA12. For the ExoFOP, this paper can be cited, or the following language can be included in the paper acknowledgments: 

\begin{displayquote}
  ``This research has made use of the Exoplanet Follow-up Observation Program (ExoFOP; DOI: 10.26134/ExoFOP5) website, which is operated by the California Institute of Technology, under contract with the National Aeronautics and Space Administration under the Exoplanet Exploration Program.''  
\end{displayquote}

Both services (NEA and ExoFOP) can also be referenced in AAS journal articles with \texttt{facilities} keywords. In addition to the above, many individual datasets hosted on the NEA and ExoFOP also request acknowledgment of their usage in publications.\footnote{Acknowledging datasets at the NEA: \url{https://exoplanetarchive.ipac.caltech.edu/docs/acknowledge.html}} We encourage users to check whether the data they are using requests acknowledgment.

\section{Conclusions} \label{sec:conclusions}

The NASA Exoplanet Archive and Exoplanet Follow-up Observing Program services provide valuable, highly utilized resources to the exoplanet community by maintaining up-to-date, comprehensive, thoroughly sourced, readily accessible, and easily explorable data describing exoplanets and their host stars. In addition to professional astronomers, the sites are used by amateur astronomers, citizen scientists, and educators in the classroom. The NEA has recently completed a large restructuring effort for our data and interfaces to simplify access to data holdings, both interactively and programmatically, and to streamline our ability to ingest and serve new data and develop new tools. This paper has outlined the current status of the data, tools, and accessibility after that reorganization, and our plans to continue improving and extending our services. The next decade promises to be the most exciting era of exoplanet discovery and characterization to date, and the objectives of the NEA and ExoFOP are to support the community during this time and help maximize exoplanet science.

\begin{acknowledgments}
The NASA Exoplanet Archive and Exoplanet Follow-up Observing Program service are operated by the California Institute of Technology, under contract with the National Aeronautics and Space Administration under the Exoplanet Exploration Program. Firefly development by Caltech/IPAC has been supported by NASA, principally through IRSA, and by the National Science Foundation, through the Vera C. Rubin Observatory.
\end{acknowledgments}

%

\vspace{5mm}


\software{Firefly \citep{joliet201}, \texttt{astropy} \citep{astropy:2013, astropy:2018, astropy:2022}, \texttt{Astroquery} \citep{Astroquery}, GNU \texttt{wget}, \texttt{Plotly Dash} \citep{dash}, \texttt{Bokeh} \citep{bokeh}, \texttt{EXOFAST} \citep{Eastman2013}}



\bibliography{sample631}{}
\bibliographystyle{aasjournal}



\end{document}